%% file: 20_01_18_ICC_CameraReady.tex
\documentclass[conference,romanappendices,10pt]{IEEEtran}

\IEEEoverridecommandlockouts

\usepackage{algorithm,algorithmic,amsmath,amssymb,amsthm,bbm,cite,color,graphicx,url}
\usepackage[USenglish]{babel}
\usepackage[utf8]{inputenc} 
\usepackage[T1]{fontenc}

{}
{}
{\newtheorem{lemma}{Lemma}}
{}
{}
{}
{}

\input{./Definitions.tex}

\newcommand{\ignore}[1]{}
\newcommand{\rmA}{\textnormal{\tiny{A1}}}
\newcommand{\rmBL}{\textnormal{\tiny{BL}}}
\newcommand{\rmUE}{\textnormal{\tiny{UE}}}
\newcommand{\rmNLoS}{\textnormal{\tiny{NLoS}}}
\newcommand{\rmLoS}{\textnormal{\tiny{LoS}}}

\title{D2D-Aided Multi-Antenna Multicasting}
\author{
\IEEEauthorblockN{Placido Mursia,$^{1}$ Italo Atzeni,$^{1}$ David Gesbert,$^{1}$ and Mari Kobayashi$^{2}$}
\IEEEauthorblockA{$^{1}$EURECOM, Communication Systems Department, Sophia Antipolis, France \\
$^{2}$Technical University of Munich, Munich, Germany \\
Emails: \{placido.mursia, italo.atzeni, david.gesbert\}@eurecom.fr, mari.kobayashi@tum.de}

\thanks{The work of P.~Mursia is supported by Marie Sk\l{}odowska-Curie actions (MSCA-ITN-ETN 722788 SPOTLIGHT). The work of I.~Atzeni, D.~Gesbert, and M.~Kobayashi is supported by the French-German Academy towards Industry 4.0 (SeCIF project) under Institut Mines-Telecom.} \vspace{-1.5mm}}

\makeindex

\begin{document}

\maketitle

\begin{abstract}
Multicast services, whereby a common valuable message needs to reach a whole population of user equipments (UEs), are gaining attention on account of new applications such as vehicular networks. As it proves challenging to guarantee decodability by every UE in a large population, service reliability is indeed the Achilles' heel of multicast transmissions. To circumvent this problem, a two-phase protocol capitalizing on device-to-device (D2D) links between UEs has been proposed, which overcomes the vanishing behavior of the multicast rate. In this paper, we revisit such a D2D-aided protocol in the new light of precoding capabilities at the base station (BS). We obtain an enhanced scheme that aims at selecting a subset of UEs who cooperate to spread the common message across the rest of the network via D2D retransmissions. With the objective of maximizing the multicast rate under some outage constraint, we propose an algorithm with provable convergence that jointly identifies the most pertinent relaying UEs and optimizes the precoding strategy at the BS.
\end{abstract}


\section{Introduction} \label{sec:INTRO}
Multicast communications, in which a transmitter wishes to convey a common message to multiple receivers, arise naturally in various wireless network scenarios. Specifically, multicast services are relevant for many challenging applications ranging from wireless edge caching, where popular contents are cached during off-peak hours and subsequently distributed \cite{Mad14,Pas16}, to the broadcasting of safety messages in vehicular networks. In parallel, device-to-device (D2D) communications have recently gained momentum on account of emerging applications such as multicasting, machine-to-machine communication, and cellular-offloading \cite{Asa14,Teh14}, and are expected to be included in the upcoming fifth-generation (5G) wireless system as a new paradigm for enhancing the network's performance \cite{Boc14,Bas14,Gup15}. 

It is well known that the multicast capacity is limited by the user equipments (UEs) in poor channel conditions and vanishes when the number of UEs increases for a fixed number of base station (BS) antennas \cite{Jin06,Sid06}. To overcome this issue, one possible approach is to let the BS focus its transmission towards a suitably selected group of UEs while leaving a few others in outage \cite{Ntr09,Meh13}. As an alternative to discarding UEs with unfavorable channel conditions, some enhanced multicasting schemes leveraging UE cooperation enabled by D2D links have been proposed to tackle the vanishing behavior of the multicast capacity when the UE population becomes large \cite{Khi06,Sir10,Hou09,Yin14,San18}. In this context, the UEs who are able to decode the common message sent by the BS act as opportunistic relays and cooperate in a subsequent phase to retransmit the information to the remaining ones. This framework, which is briefly reviewed in Section~\ref{sec:MO}, has been considered so far only for the case of single-antenna BS.

In this paper, we point out that endowing the BS with precoding capabilities radically transforms the above scheme both in nature and performance, since we can leverage the precoding gain brought by multiple antennas in addition to the D2D links. However, this implies the joint optimization of the multicast rate and precoding strategy at the BS, which is highly complex to tackle: to the best of our knowledge, this is the first work that addresses such a scenario. Specifically, we propose a two-phase scheme where the BS targets a strategic subset of UEs in the first phase (to be optimized) via suitable precoding, while in the second phase the UEs who have successfully decoded the common message simultaneously retransmit it to the remaining ones. In addition, this paper includes a notion of {\em target outage} in the multicast service, by which the multicast rate is the encoding rate of the common message which guarantees decodability by  most UEs while strategically avoiding to waste resources on a small amount of UEs with particularly unfavorable channel conditions \cite{Des18}. In this respect, we propose an iterative, low-complexity algorithm that seeks to find both a suitable subset of UEs to be targeted in the first phase and the maximum multicast rate that can be supported by the system. The proposed algorithm is shown to converge to a locally optimal solution. Numerical results reveal its superior performance with respect to a benchmark single-phase scheme where the UEs are served by means of precoded multicast only.

\section{System Model} \label{sec:SM}
Consider a wireless network where a BS equipped with $M$ transmit antennas aims at conveying a common message to a set $\setK \triangleq \{1, \ldots, K\}$ of single-antenna UEs. The UEs are also connected to each other via D2D links and operate in half-duplex mode. We focus on a two-phase scheme similar to the one proposed in \cite{Khi06,San18,Sir10}, where the BS multicasts the common message in the first phase and the UEs who have successfully decoded jointly retransmit the information in the second phase through the D2D links. In this context, the BS can cleverly use its resources by transmitting to a suitably\addtolength{\voffset}{0.635mm} selected subset of UEs, which in turn cooperate to spread the common message across the rest of the network. We use $\h_{k} \in \Compl^{M \times 1}$ and $h_{k j} \in \Compl$ to denote the channels from the BS to UE~$k$ and from UE~$j$ to UE~$k$, respectively. Assuming that the channels remain constant for each phase, the considered two-phase scheme is described as follows.
\begin{itemize}
\item[\textit{i)}] During the multicast transmission in the first phase, the BS transmits the common message at a rate $r$, referred to as \textit{multicast rate} (to be optimized). The receive signal at UE~$k$ in the first phase is given by
\begin{align}\label{eq:y_k}
y_{k,1} \triangleq \sqrt{p} \h_k^{\herm}\x + n_k, \qquad \forall k \in \setK
\end{align}
where $p$ denotes the transmit power at the BS, $\x \in \Compl^{M\times 1}$ is the transmit signal with transmit covariance matrix $\Sigmab \triangleq \Exp[\x\x^{\herm}] \in \Compl^{M\times M}$ (to be optimized), and $n_k \sim \mathcal{CN}(0,\sigma^2)$ is the noise term at UE~$k$. The common message is decoded by UE~$k$ if its achievable rate in the first phase is at least $r$, i.e., if $\log_2(1+ \rho \h_k^{\herm} \Sigmab \h_k) \geq r$, where $\rho \triangleq p/\sigma^{2}$ denotes the transmit signal-to-noise ratio (SNR) at the BS in the first phase. We thus define $\setU \triangleq \big\{ k \in \setK : \log_2(1+ \rho \h_k^{\herm} \Sigmab \h_k) \geq r \big\}$ as the subset of UEs who decode successfully in the first phase and $\setU^{\prime} \triangleq \setK \setminus \setU$.
\item[\textit{ii)}] During the cooperative D2D transmission in the second phase, the common message is simultaneously retransmitted by the UEs in $\setU$ in an isotropic fashion such that the UEs in $\setU^{\prime}$ receive a non-coherent sum of the D2D transmit signals. The receive signal at UE~$k$ in the second phase is given by
\begin{align}\label{eq:y_kj}
y_{k,2} & \triangleq \sum_{j \in \setU} \sqrt{p_j} h_{kj} x_j + n_k, \qquad \forall k \in \setU^{\prime}
\end{align}
where $p_j$ denotes the transmit power at UE~$j$ and $x_j \in \Compl$ represents the transmit signal with $\Exp[|x_j|^2]=1$.\footnote{Note that the UEs transmit with fixed power and do not perform any power control in the second phase.} The common message is decoded by UE~$k$ if its achievable rate in the second phase is at least $r$, i.e., if $\log_2 \big( 1+ \big| \sum_{j \in \setU} \sqrt{\rho_{j}} h_{k j} \big|^{2} \big) \geq r$, where $\rho_j \triangleq p_j/\sigma^2$ denotes the transmit SNR at UE~$j$ in the second phase.
\end{itemize}

\section{Multicasting Optimization} \label{sec:MO}
In this section, we first briefly recall some existing results on the multicast capacity as well as some recently proposed enhanced multicasting schemes.

Given the receive signal in \eqref{eq:y_k}, the multicast capacity is given by \cite{Jin06}
\begin{align} 
C(\H) & \triangleq \max_{\Sigmab \succeq \0 \; : \; \tr(\Sigmab)\leq 1} \; \min_{k \in \setK} \log_2(1+ \rho \h_k^{\herm} \Sigmab \h_k) \label{eq:mult_capac} \\
&= \log_2\left(1+ \rho \max_{\Sigmab \succeq \0 \; : \; \tr(\Sigmab)\leq 1} \; \min_{k \in \setK}  \h_k^{\herm} \Sigmab \h_k\right) \label{eq:mult_capac_2}
\end{align}
\noindent where $\H \triangleq [\h_1, \ldots ,\h_K] \in \Compl^{M \times K}$. Observe that problems~\eqref{eq:mult_capac}--\eqref{eq:mult_capac_2} are convex and, although no closed-form solution is available, the capacity-achieving transmit covariance matrix can be efficiently obtained by means of semidefinite programming techniques. Evidently, the multicast capacity is limited by the UEs in poor channel conditions: in particular, when the number of UEs $K$ grows large and the number of antennas $M$ remains fixed, $C(\H)$ scales as $1/K^{1/M}$ for the case of i.i.d Rayleigh fading channel \cite{Jin06}. Enhanced multicasting schemes have been proposed to overcome such vanishing behavior of the multicast capacity when $K$ increases. Particularly relevant to this paper are \textit{multicasting with UE selection} and \textit{D2D-aided multicasting}.

\textit{a) Multicasting with UE selection.} Only a subset of UEs with favorable channel conditions is ensured to decode the common message, while the other UEs are allowed not to decode. For i.i.d. Rayleigh fading channels, if $M$ and $K$ grow large with a constant ratio, the multicast rate of $(1-\nu) \log M$ can be achieved by ensuring that a subset of $M^{\nu}$ randomly selected UEs decode the common message \cite{Jin06}.

\textit{b) D2D-aided multicasting.} Instead of discarding UEs in poor channel conditions, D2D-aided multicasting operates in two phases so that the UEs who have decoded successfully in the first phase jointly retransmit the common message to the other UEs in the second phase. For the single-antenna case with i.i.d. Rayleigh fading channels, \cite{Khi06} proposes a simple two-phase protocol and proves that, when $K$ grows asymptotically large, the optimal average multicast rate is $\log_{2}(1+Q/\sigma^{2})$, where $Q$ is the aggregate power consumed by the network over the two phases. The performance of a two-phase cooperative multicasting scheme is studied in a more general network topology in \cite{Sir10,San18}. Both a dense model (with a finite network area) and an extended model (with a finite density) are considered. 
A similar two-phase multicasting scheme has been also studied by taking into account fairness between users or energy efficiency (see, e.g., \cite{Hou09,Yin14} and references therein). 

\section{D2D-Aided Multi-Antenna Multicasting} \label{sec:D2D-MAM}
This paper considers a D2D-aided multicasting scenario where the BS is equipped with multiple antennas. The anticipated advantage of multi-antenna BS in this context is that precoding allows to spatially target a subset of UEs who are strategically located with respect to both the BS and the remaining UEs. To the best of our knowledge, this is the first work to tackle such a scenario.

Assume that the BS transmits at a multicast rate $r$ with transmit covariance matrix $\Sigmab$. Let us define the binary variables
\begin{align}
\label{eq:Z_k1} Z_{k,1}(r, \Sigmab) & \! \triangleq \! \mathbbm{1} \big[ \log_{2} (1 + \rho \h_{k}^{\herm} \Sigmab \h_{k}) \geq r \big], \\
\label{eq:Z_k2} Z_{k,2}(r, \Sigmab) & \! \triangleq \! \mathbbm{1} \bigg[ \log_{2} \! \bigg( 1 \! + \! \bigg| \sum_{j \in \setK \setminus \{ k \}} \! Z_{j,1}(r,\Sigmab) \sqrt{\rho_{j}} h_{k j} \bigg|^2 \bigg) \! \geq r \bigg]
\end{align}
where $Z_{k,i}$ is equal to $1$ if UE~$k$ decodes successfully in the $i$th phase and to $0$ otherwise. Accordingly, the probabilities\addtolength{\voffset}{0.635mm} that UE~$k$ decodes successfully in the first or second phase are given by
\begin{align}
\label{eq:P_k1} P_{k,1}(r, \Sigmab) & \triangleq \Pr \big[ Z_{k,1}(r, \Sigmab) = 1 \big], \\
\label{eq:P_k2} P_{k,2}(r, \Sigmab) & \triangleq \Pr \big[ Z_{k,2}(r, \Sigmab) = 1 \big]
\end{align}
respectively. The probability that UE~$k$ decodes successfully over the two phases is thus obtained as
\begin{align}
P_{k}(r, \Sigmab) & \triangleq P_{k,1}(r, \Sigmab) + \big( 1-P_{k,1}(r, \Sigmab) \big) P_{k,2}(r, \Sigmab)
\end{align}
and we define the \textit{average success probability} as
\begin{align} \label{eq:av_P}
\bar{P}(r, \Sigmab) \triangleq \frac{1}{K} \sum_{k \in \setK} P_{k}(r, \Sigmab)
\end{align}
which denotes the probability that a randomly chosen UE decodes successfully over the two phases. In this context, we assume that the time resource is equally divided between the two phases. Hence, the \textit{outage multicast rate}, i.e., the maximum transmission rate at which a randomly chosen UE decodes successfully with probability at least $1-\epsilon$ over the two phases, is defined as
\begin{align} \label{eq:out_R}
\tilde{R}(r,\Sigmab) \triangleq \frac{r}{2} \quad \textrm{with $r$ solution to $\bar{P}(r, \Sigmab)\geq 1-\epsilon$}
\end{align}
with $\epsilon \in [0,1)$ being the target outage. This parameter describes the trade-off between reliability and multicast rate. Namely, a low value of $\epsilon$ forces the system to serve the UEs with high reliability while decreasing the multicast rate. 

\subsection{Problem Formulation} \label{sec:D2D-MAM_prob}
Our objective is to jointly optimize the multicast rate $r$ and the transmit covariance matrix $\Sigmab$ that maximize the outage multicast rate over the two phases in \eqref{eq:out_R}, i.e.,
\begin{align} \label{eq:prob}
\begin{array}{cl}
\displaystyle \max_{r>0, \; \Sigmab \succeq \0} & r \\
\mathrm{s.t.}& \mathrm{tr}(\Sigmab) \leq 1, \\
& \bar{P}(r, \Sigmab) \geq 1-\epsilon
\end{array}
\end{align}
with $\bar{P}(r, \Sigmab)$ defined in \eqref{eq:av_P}. In this paper, we consider the case of perfect channel state information (CSI), where the BS has perfect knowledge of the downlink channels $\{ \h_{k} \}_{k \in \setK}$ and D2D channels $\{ h_{k j} \}_{k,j \in \setK}$, whereas each UE $k$ has perfect knowledge of the effective downlink channel power gain $\h_{k}^{\herm}\Sigmab\h_k$ and the D2D channels. The more general case of imperfect/partial CSI is left for future work. In this context, the probabilities that UE~$k$ decodes successfully in the first or second phase introduced in \eqref{eq:P_k1}--\eqref{eq:P_k2} become
\begin{align}
\label{eq:P_k1_perfCSI} P_{k,1}(r, \Sigmab) & = Z_{k,1}(r, \Sigmab), \\
\label{eq:P_k2_perfCSI} P_{k,2}(r, \Sigmab) & = Z_{k,2}(r, \Sigmab)
\end{align}
respectively, and the average success probability in \eqref{eq:av_P} may be rewritten as
\begin{align}
\bar{P}(r, \Sigmab) \! = \! \frac{1}{K} \sum_{k \in \setK} \big( Z_{k,1}(r, \Sigmab) \! + \! \big( 1 \! - \! Z_{k,1}(r, \Sigmab) \big) Z_{k,2}(r, \Sigmab) \big).
\end{align}
\addtolength{\voffset}{0.635mm}
\subsection{Baseline Algorithm} \label{sec:D2D-MAM_bl}
Assuming that the whole data transmission occurs in the first phase, problem \eqref{eq:prob} is solved by selecting the UEs served by the BS and optimizing the transmit covariance matrix that maximize the outage multicast rate. Note that, in this case, the outage constraint in \eqref{eq:prob} is expressed as $\sum_{k \in \setK}P_{k,1}(r, \Sigmab)/K\geq 1-\epsilon$.\footnote{Without loss of generality, we assume that $\epsilon$ is chosen such that $(1-\epsilon) K$ is an integer number.} However, the problem of deriving the optimal UE selection strategy in the first phase is NP-hard, as it requires to evaluate all possible combinations of $(1-\epsilon)K$ UEs. As a consequence, we build on the intuition below to derive a suboptimal UE selection scheme, referred to as \textit{baseline algorithm}. Observe that, while its problem formulation is also novel, the baseline algorithm serves as a benchmark to demonstrate the gains obtained by allowing a second phase of cooperative D2D retransmissions.
 
\begin{lemma} \label{lem:baseline}
For a class of channels satisfying $\Exp[\h_k \h_k^{\herm} ] = \gamma_k \I_M$ for $\{\gamma_k > 0\}_{k \in \setK}$, the optimal UE selection strategy with statistical channel knowledge is the one choosing the $(1-\epsilon)K$ UEs with the highest $\gamma_k$.
\end{lemma}

\begin{IEEEproof}
If $\{\gamma_k > 0\}_{k \in \setK}$ are known at the BS, we have
\begin{align}
\nonumber & \hspace{-2mm} \max_{\setU \subset \setK \; : \; |\setU | = (1 - \epsilon) K} \Exp \bigg[ \max_{\Sigmab \succeq \0 \; : \; \tr(\Sigmab) \leq 1} \; \min_{k \in \setU} \h_k^{\herm} \Sigmab \h_k \bigg] \\
\label{eq:lem1_proof1} & \hspace{-1mm} \leq \max_{\setU \subset \setK \; : \; |\setU| = (1 - \epsilon) K} \; \max_{\Sigmab \succeq \0 \; : \; \tr(\Sigmab) \leq 1} \; \min_{k \in \setU} \Exp[\h_k^{\herm} \Sigmab \h_k ] \\
& \hspace{-1mm} = \max_{\setU \subset \setK \; : \; |\setU|=(1-\epsilon)K} \; \max_{\Sigmab \succeq \0 \; : \; \tr(\Sigmab) \leq 1} \; \min_{k \in \setU} \tr \big( \Sigmab \Exp[\h_k \h_k^{\herm}] \big) \\
\label{eq:lem1_proof2} & \hspace{-1mm} = \max_{\setU \subset \setK \; : \; |\setU| = (1 - \epsilon) K} \; \min_{k \in \setU} \gamma_k \vspace{-1mm}
\end{align}
where \eqref{eq:lem1_proof1} follows from the concavity of the function $\min_{k \in \setU} \h_k^{\herm} \Sigmab \h_k$ and \eqref{eq:lem1_proof2} is due to the fact that the optimal $\Sigmab$ satisfies $\tr(\Sigmab) = 1$. Finally, it follows that \eqref{eq:lem1_proof2} is given by the solution presented in the lemma. \vspace{1mm}
\end{IEEEproof}

\noindent Lemma~\ref{lem:baseline} states that, if the channels have a structure such that the UEs can be ordered statistically, the exhaustive search over all possible sets reduces to a simple selection of the $(1-\epsilon) K$ UEs with highest average channel power gain $\gamma_{k}$. Motivated by this observation, we propose to adapt such a UE selection to the case of perfect CSI at the BS. More precisely, we build the subset $\setU$ by selecting the $(1-\epsilon) K$ UEs with the highest channel power gain $\| \h_k \|^2$. Then, we compute the transmit covariance matrix that achieves the multicast capacity over the subset $\setU$, i.e.,
\begin{align}\label{eq:Sigma_bsl}
\Sigmab_{\rmBL} = \argmax_{\Sigmab \succeq \mathbf{0} \; : \; \mathrm{tr}(\Sigmab) \leq 1} \min_{k\in \setU} \h_{k}^{\herm}\Sigmab\h_{k}.
\end{align}
Since the whole time resource is dedicated to the first phase, the resulting outage multicast rate is given by
\begin{align}\label{eq:r_bsl}
r_{\rmBL} = \log_2\left(1 + \rho \min_{k\in \setU} \h_{k}^{\herm}\Sigmab_{\rmBL}\h_{k}\right).
\end{align}


%

\subsection{D2D-MAM Algorithm} \label{sec:D2D-MAM_alg}
By adding a second phase of cooperative D2D retransmissions, the problem of jointly optimizing the transmit covariance matrix and the multicast rate becomes consistently more difficult to tackle. Therefore, we resort to a heuristic approach. In this respect, we propose an efficient iterative algorithm based on the alternating optimization of the transmit covariance matrix $\Sigmab$ and the multicast rate $r$. The goal is to serve a suitably selected subset of UEs in the first phase by means of precoding at the BS such that the outage multicast rate is maximized.

At each iteration $n$, the algorithm computes the transmit covariance matrix $\Sigmab^{(n)}$ that achieves the multicast capacity over a predetermined subset of UEs $\setU^{(n-1)}$. Then, the multicast rate $r^{(n)}$ is computed as the maximum rate that guarantees the outage constraint over the two phases given the transmit covariance matrix obtained in the previous step, i.e., $\bar{P}(r^{(n)}, \Sigmab^{(n)}) \geq 1 - \epsilon$. In turn, $r^{(n)}$ yields an updated subset $\setU^{(n)}$ of UEs decoding successfully in the first phase, and a new transmit covariance matrix is obtained by optimizing over $\setU^{(n)}$. This procedure is iterated until the multicast rate converges. The proposed algorithm is referred to in the following as \textit{D2D-aided multi-antenna multicasting (D2D-MAM) algorithm} and is formally described in Algorithm~\ref{alg:A1}. Despite being suboptimal, the D2D-MAM~algorithm has the key advantage of not requiring any tuning parameter selection. Furthermore, it converges to a local optimum of problem~\eqref{eq:prob}, as formalized in the following lemma.

\begin{lemma}\label{lem:alg1}
The D2D-MAM~algorithm converges to a local optimum of problem~\eqref{eq:prob}.
\end{lemma}

\begin{IEEEproof}
Since step~(S.1) of Algorithm~\ref{alg:A1} optimizes $\Sigmab^{(n)}_{\rmA}$ over $\setU^{(n-1)}$, we have
\begin{align} \label{eq:Sigma_n}
\min_{k\in \setU^{(n-1)}} \h_k^{\herm}\Sigmab^{(n)}_{\rmA}\h_k \geq \min_{k\in \setU^{(n-1)}} \h_k^{\herm}\Sigmab^{(n-1)}_{\rmA}\h_k
\end{align}
i.e., the minimum rate achievable by the UEs in $\setU^{(n-1)}$ increases with the new transmit covariance matrix $\Sigmab^{(n)}_{\rmA}$. Furthermore, at each iteration $n$ of the D2D-MAM~algorithm, the following holds:
\begin{align}
r^{(n)}_{\rmA} & \geq \log_2 \bigg(1 + \rho \min_{k\in \setU^{(n-1)}} \h_k^{\herm}\Sigmab^{(n)}_{\rmA}\h_k \bigg) \label{eq:r_n1} \\
& \geq \log_2 \bigg( 1 + \rho \min_{k\in \setU^{(n-1)}} \h_k^{\herm}\Sigmab^{(n-1)}_{\rmA}\h_k \bigg) \label{eq:r_n2} \\
& \geq r^{(n-1)}_{\rmA} \label{eq:r_n3}
\end{align}
where \eqref{eq:r_n1} follows from step~(S.2) of Algorithm~\ref{alg:A1} (by which it is possible to increase the multicast rate as long as the outage constraint is guaranteed), \eqref{eq:r_n2} is a direct consequence of \eqref{eq:Sigma_n}, and \eqref{eq:r_n3} follows from the fact that $\setU^{(n-1)}$ contains the UEs whose achievable rate in the first phase is at least $r^{(n-1)}_{\rmA}$. Hence, the multicast rate cannot decrease between consecutive iterations. Finally, if $\setU^{(n)}=\setU^{(n-1)}$, then it is not possible to further increase the multicast rate and $r^{(n)}=r^{(n-1)}$, which implies that convergence is reached.
\end{IEEEproof} \vspace{1mm}

Regarding the optimization of the multicast rate in step~(S.2) of Algorithm~\ref{alg:A1}, we have
\begin{align}
r_{\rmA}^{(n)} \in \bigg[ r_{\rmA}^{(n-1)}, \log_2 \bigg( 1+\rho \max_{k \in \setU^{(n-1)}} \h_k^{\herm}\Sigmab_{\rmA}^{(n)}\h_k \bigg) \bigg]
\end{align}
where the lower bound follows from Lemma~\ref{lem:alg1} and the upper bound is necessary to guarantee that at least one UE is served in the first phase. Therefore, $r_{\rmA}^{(n)}$ can be computed by means of linear search over the above interval. Accordingly, every iteration of the D2D-MAM algorithm requires the solution of a convex problem in step~(S.1) and a linear search in step~(S.2). This scheme thus provides a considerable complexity reduction with respect to problem~\eqref{eq:prob}, which is non convex in both $r$ and $\Sigmab$.
\begin{figure}[t!]
\vspace{-3mm}
\begin{algorithm}[H] \caption{(D2D-MAM)} \label{alg:A1}
\smallskip
\begin{algorithmic}
\STATE \texttt{\hspace{-0.8mm} Data \hspace{-1.3mm} :} Fix $\setU^{(0)} = \setK$ and $n=1$.
\STATE \texttt{(S.1) :} Optimize the transmit covariance matrix as
\begin{align*}
\Sigmab^{(n)}_{\rmA} = \argmax_{\Sigmab \succeq \mathbf{0} \; : \; \mathrm{tr}(\Sigmab) \leq 1} \min_{k \in \setU^{(n-1)}} \h_{k}^{\herm}\Sigmab\h_{k}.
\end{align*}
\STATE \texttt{(S.2) :} Maximize the multicast rate as \\
\hspace{1.46cm}		$r^{(n)}_{\rmA}= \max\{r:  \bar{P}(r,\Sigmab^{(n)}_{\rmA}) = 1 - \epsilon\}$. 
\STATE \texttt{(S.3) :} Update the subset of UEs decoding successfully $\phantom{////~\qquad}$ in the first phase as
\begin{align*}
\setU^{(n)} = \big\{ k \, : \, \log_{2} (1 + \rho \h_{k}^{\herm} \Sigmab^{(n)}_{\rmA} \h_{k}) \geq r^{(n)}_{\rmA} \big\}.
\end{align*}
\STATE \texttt{(S.4) :} \texttt{If} $r_{\rmA}^{(n)} = r_{\rmA}^{(n-1)}$: fix $\Sigmab_{\rmA} = \Sigmab^{(n)}_{\rmA}$ and \\
\hspace{1.46cm} 		 $r_{\rmA} = r_{\rmA}^{(n)}$; \texttt{Stop}. \\
\hspace{1.46cm} 		 \texttt{Else}: $n \leftarrow n+1$; \texttt{Go to (S.1)}.
\end{algorithmic}
\end{algorithm}
\vspace{-5mm}
\end{figure}

\section{Numerical Results and Discussion} \label{sec:NR}
In this section, we present numerical results to analyze the benefits of the two-phase scheme proposed in Section~\ref{sec:D2D-MAM_alg} with respect to the single-phase baseline scheme described in Section~\ref{sec:D2D-MAM_bl}.

The channel model used for our numerical results is described as follows. Let $\h_{k} =  \sqrt{\gamma_{k}} \eta_{k} \a(\theta_{k})$ and $h_{k j} =  \sqrt{\gamma_{k j}} \eta_{k j}$, where $\gamma_{k}$ and $\gamma_{k j}$ are the average channel power gains, $\eta_{k}$ and $\eta_{k j}$ are the small-scale fading coefficients, and $\a(\theta_{k}) \triangleq [1 \ e^{-i2\pi\delta\cos(\theta_k)} \ldots e^{-i2\pi\delta(M-1)\cos(\theta_k)}]^{\tran} \in \Compl^{M\times 1}$ is the linear array response vector at the BS for the steering angle $\theta_k$, with $\delta = 0.5$ being the ratio between the antenna spacing and the signal wavelength. In particular, we set $\gamma_k = \beta d_{k}^{-\alpha_k}$ and $\gamma_{kj} = \beta d_{kj}^{-\alpha_{kj}}$, where $\beta$ is the average channel power gain at a reference distance, $d_k$ and $d_{kj}$ denote the distances from the BS to UE~$k$ and from UE~$j$ to UE~$k$, respectively, and $\alpha_k$ and $\alpha_{kj}$ represent the associated pathloss exponents. We assume that $K_{\rmNLoS}$ randomly chosen UEs (out of $K$) are in non-line-of-sight (NLoS) conditions with respect to the BS. In this regard, $\alpha_{\rmNLoS}$ and $\alpha_{\rmLoS}$ denote the pathloss exponents of the UEs in NLoS and line-of-sight (LoS) conditions, respectively; moreover, we assume LoS D2D links, i.e., $\{ \alpha_{kj} = \alpha_{\rmLoS} \}_{k,j \in \setK}$. For simplicity, we assume that all UEs have the same transmit power and, thus, the same transmit SNR, i.e., $\{ \rho_j = \rho_{\rmUE} \}_{j \in \setK}$. Unless otherwise stated, we consider $M=16$, $K_{\rmNLoS} = K/2$, $\alpha_{\rmNLoS}=4$, $\alpha_{\rmLoS}=2$, $\rho=30$~dB, and $\rho_{\rmUE} = 20$~dB. Lastly, we perform our simulations by averaging over $2 \times 10^{3}$ uniformly random UE locations within a semicircular area of radius $d_{\mathrm{max}}=50$~m from the BS.

\begin{figure}[t!]
\centering \vspace{0.6mm}
\includegraphics[scale=0.8]{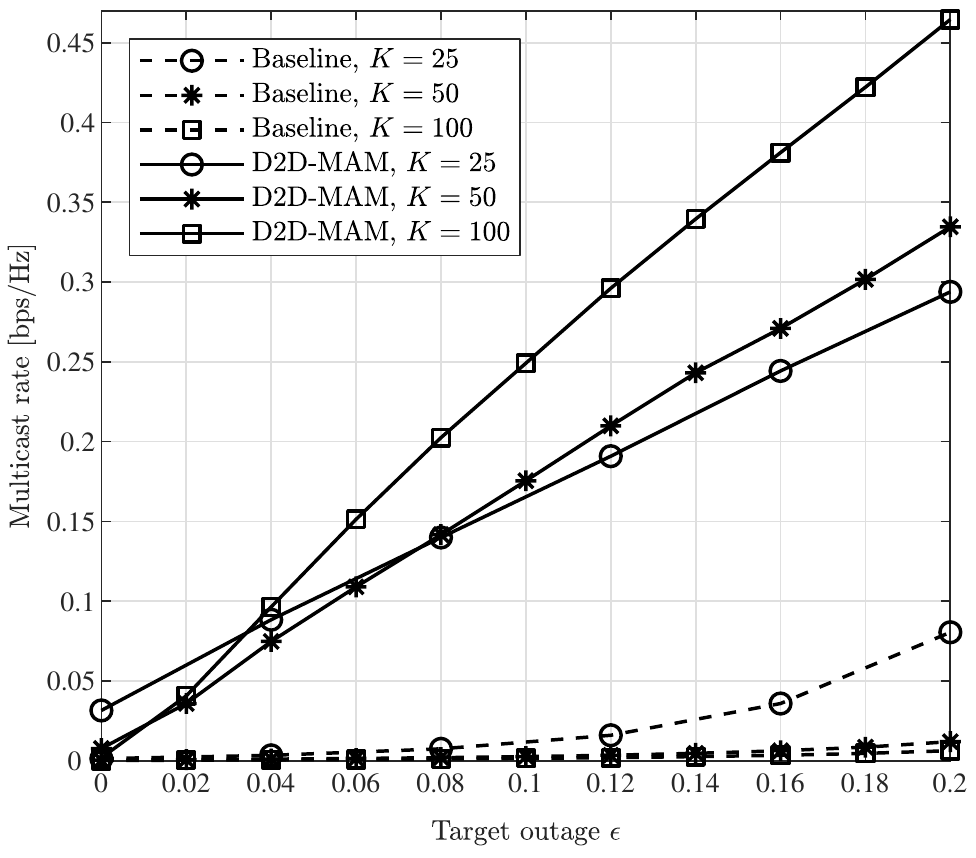}
\caption{Outage multicast rate obtained with the baseline and D2D-MAM algorithms versus the target outage $\epsilon$, with $M=16$, $K_{\rmNLoS}/K=0.5$, $\rho=30$~dB, $\rho_{\rmUE}=20$~dB, and for different values of $K$.}
\label{fig:K_vs_eps}
\end{figure}

\begin{figure}[t!]
\centering
\includegraphics[scale=0.8]{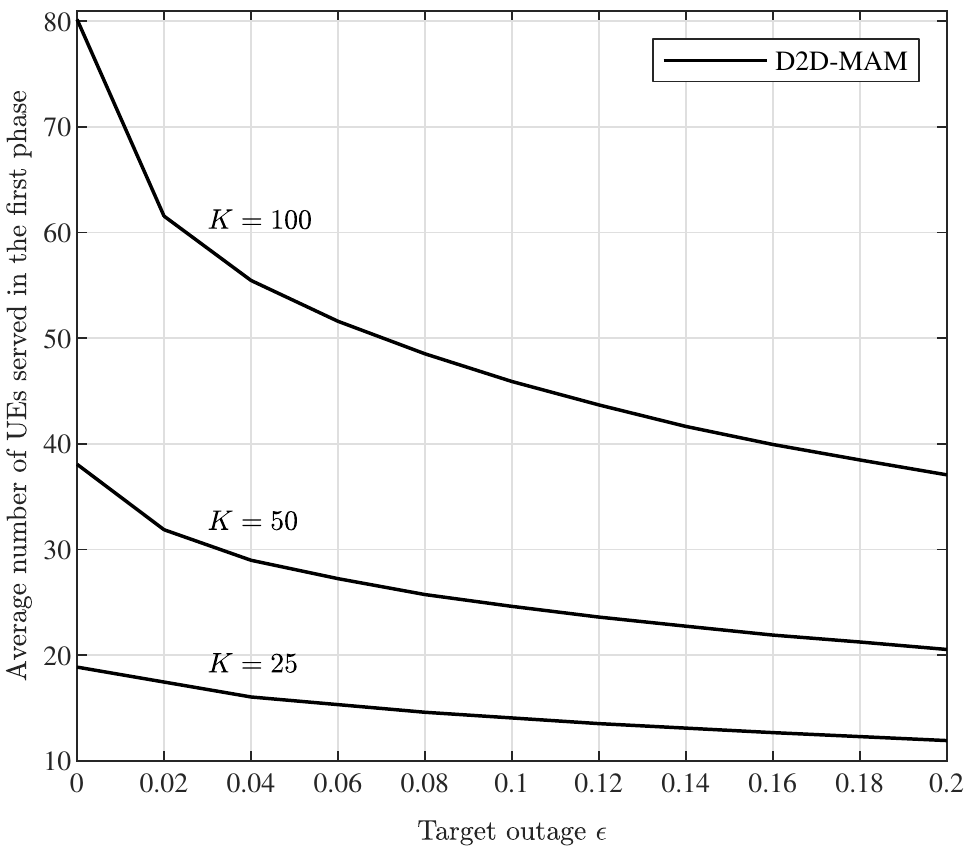}
\caption{Average number of UEs served in the first phase for the D2D-MAM algorithm versus the target outage $\epsilon$, with $M=16$, $K_{\rmNLoS}/K=0.5$, $\rho=30$~dB, $\rho_{\rmUE}=20$~dB, and for different values of $K$.}
\label{fig:Nue_vs_eps}
\end{figure}

\begin{figure}[t!]
\centering
\includegraphics[scale=0.8]{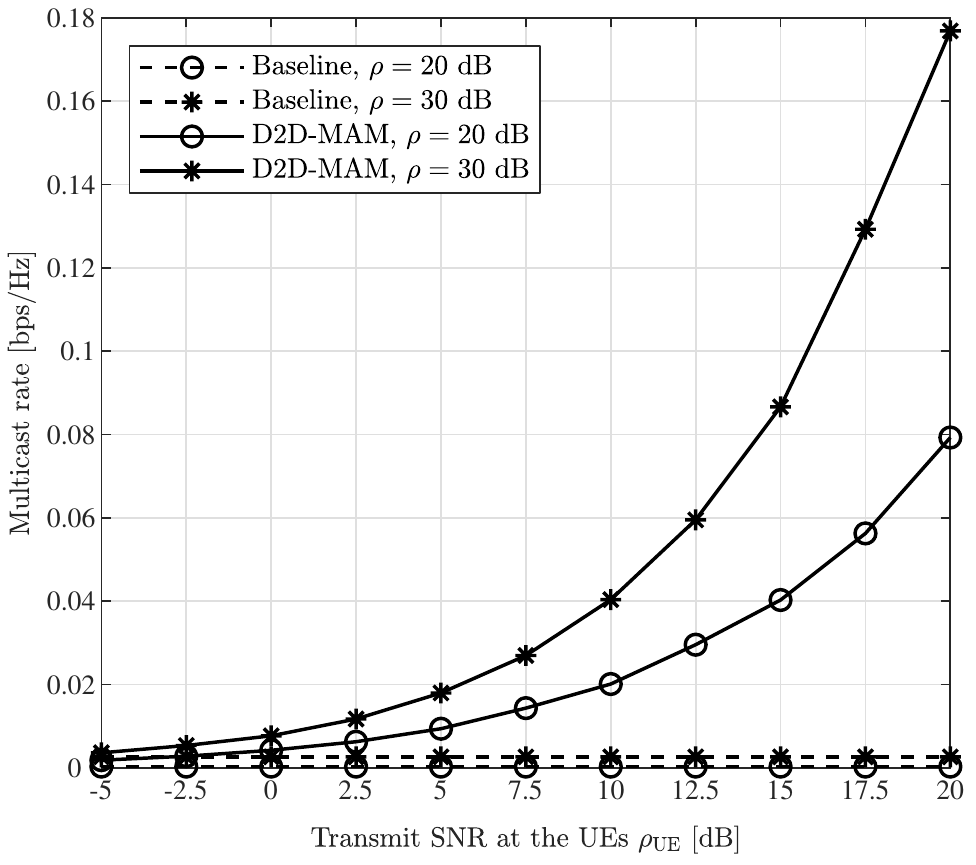}
\caption{Outage multicast rate obtained with the baseline and D2D-MAM algorithms versus the transmit SNR at the UEs $\rho_{\rmUE}$, with $M=16$, $K=50$, $K_{\rmNLoS}/K=0.5$, $\epsilon=0.1$, and for different values of $\rho$.}
\label{fig:rho_vs_rhoj}
\end{figure}

Figure~\ref{fig:K_vs_eps} illustrates the outage multicast rate for the baseline and D2D-MAM algorithms versus the target outage $\epsilon$ for different values of $K$; here, the second phase of the D2D-MAM algorithm is shown to dramatically increase the performance as compared with the single-phase baseline algorithm. Moreover, for small-to-moderate values of $\epsilon$, the outage multicast rate increases with the number of UEs (contrary to the baseline). Our proposed algorithm is thus able to enhance the performance of single-phase multicasting, which is limited by the channel of the worst UE. The corresponding average number of UEs served in the first phase for each value of $K$ is depicted in Figure~\ref{fig:Nue_vs_eps}. Remarkably, in the considered scenario, the D2D-MAM algorithm converges after very few iterations (typically between $3$ and $10$) even for large values of $K$. The benefits brought by the second phase of cooperative D2D retrasmissions are more evident in Figure~\ref{fig:rho_vs_rhoj}, where the outage multicast rate against the transmit SNR at the UEs $\rho_{\rmUE}$ is shown for different values of the transmit SNR at the BS $\rho$. Here, increasing the transmit SNRs at both the BS and the UEs has an evident impact on the performance of the D2D-MAM algorithm; on the other hand, increasing the transmit SNR at the BS does not bring substantial gains for the baseline algorithm due to the lack of spatial degrees of freedom to serve all the selected UEs. Additionally, we observe that the performance of the baseline algorithm approaches the one of the proposed D2D-MAM algorithm only when the UEs transmit with very low power in the second phase.

\begin{figure}[t!]
\centering
\includegraphics[scale=0.8]{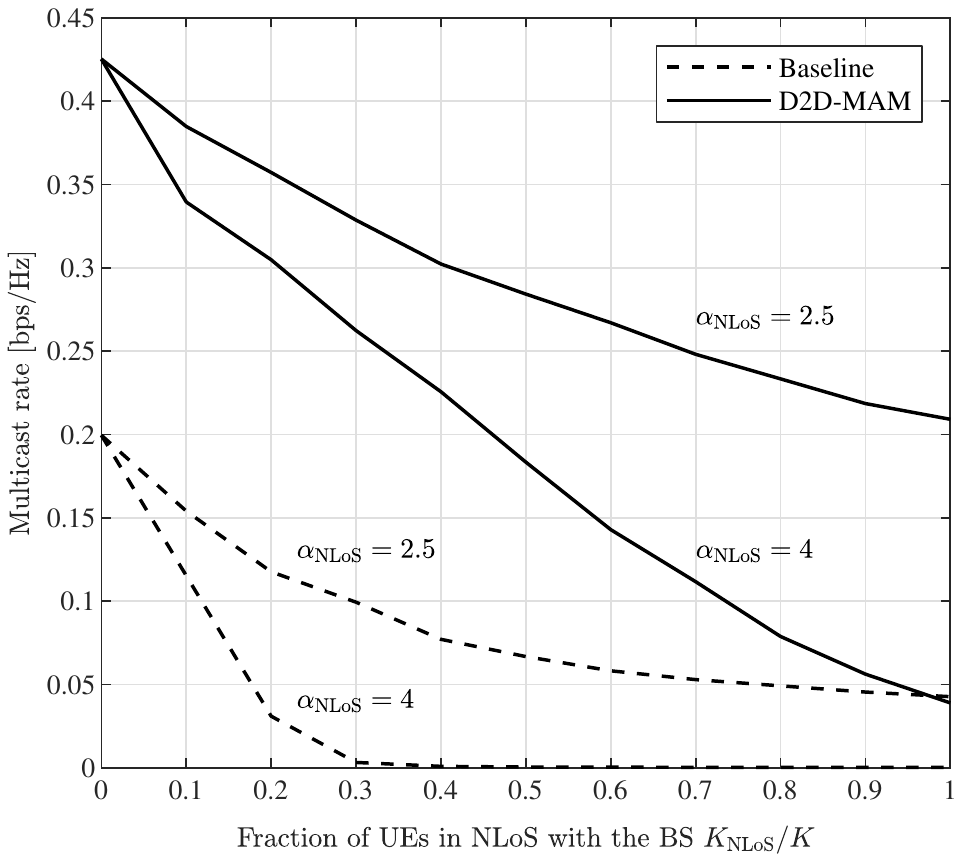}
\caption{Outage multicast rate obtained with the baseline and D2D-MAM algorithms versus the fraction of UEs in NLoS conditions $K_{\rmNLoS}/K$, with $M=16$, $K=50$, $\epsilon=0.1$, $\rho=30$~dB, $\rho_{\rmUE}=20$~dB, and for different values of $\alpha_{\rmNLoS}$.}
\label{fig:NLOS_vs_KNLOS}
\end{figure}

\begin{figure}[t!]
\vspace{0.8mm}
\centering
\includegraphics[scale=0.8]{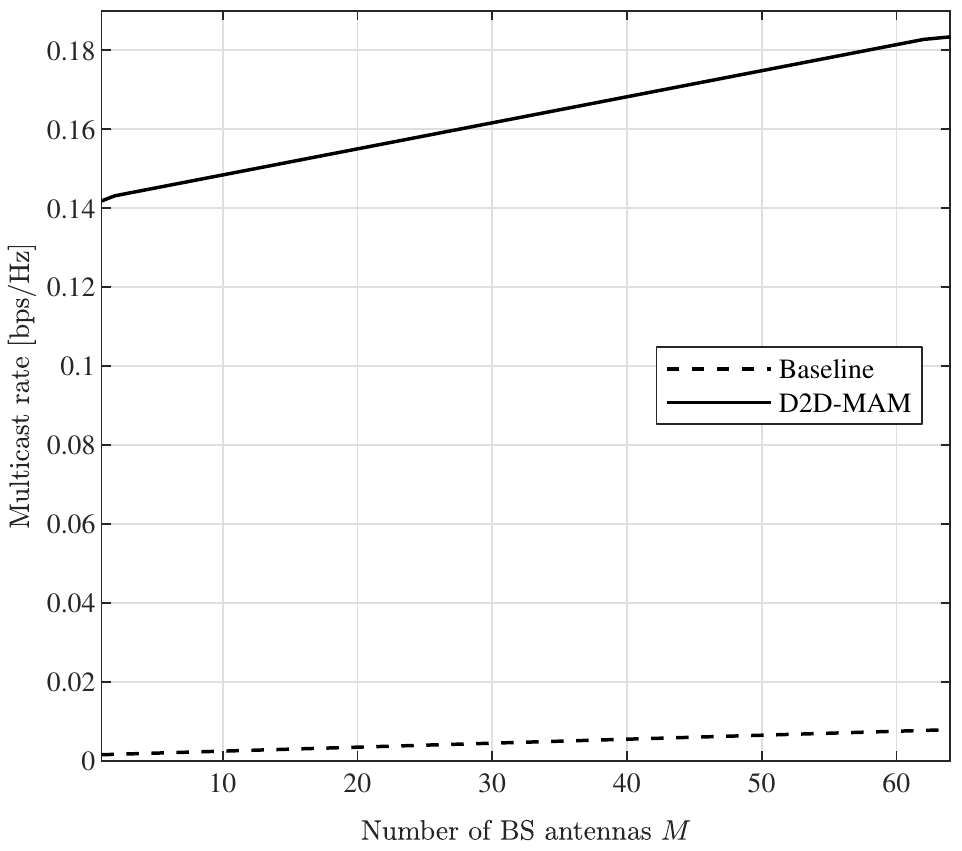}
\caption{Outage multicast rate obtained with the baseline and D2D-MAM algorithms versus the number of BS antennas $M$, with $K=50$, $K_{\rmNLoS}/K=0.5$, $\epsilon=0.1$, $\rho=30$~dB and $\rho_{\rmUE}=20$~dB.}
\label{fig:M_vs_eps}
\end{figure}

Another limiting factor in the performance of the baseline algorithm is represented by the UEs in NLoS conditions with respect to the BS. In this respect, the precoding strategy naturally conveys more power in the directions of these UEs to boost their achievable rates. Our proposed algorithm overcomes such limitation thanks to the second phase of cooperative D2D retrasmissions. This can be observed in Figure~\ref{fig:NLOS_vs_KNLOS}, where the outage multicast rate is plotted against the fraction of NLoS UEs $K_{\rmNLoS}/K$. Indeed, the D2D-MAM algorithm relies on D2D retransmissions in the second phase to reach the UEs in very poor channel conditions with respect to the BS. Lastly, Figure~\ref{fig:M_vs_eps} shows the gains brought by precoding in the first phase. In particular, as the number of BS antennas increases, the BS is able to focus its transmit power more efficiently towards the subset of UEs targeted in the first phase, which results in an overall improved outage multicast rate. Note that the lowest multicast rate is obtained for $M=1$, which corresponds to the case where the BS does not have any precoding capability.

\section{Conclusions} \label{sec:CON}
This paper considers a two-phase multicasting scheme where a BS equipped with multiple antennas transmits a common message to a suitably selected subset of UEs, which in turn cooperate to spread the information across the rest of the network via simultaneous D2D retransmissions. Leveraging the precoding gain at the multi-antenna BS and the the D2D links between UEs, we target the joint optimization of the multicast rate and precoding strategy at the BS under some outage constraint. In this respect, we propose an iterative, low-complexity algorithm that is proved to converge to a locally optimal solution. Numerical results corroborate the superior performance of our proposed algorithm with respect to a benchmark single-phase scheme where the UEs are served by means of precoded multicast only. \vspace{2mm}

\addcontentsline{toc}{chapter}{References}
\bibliographystyle{IEEEtran}
\bibliography{IEEEabrv,refs}

\end{document}

%% file: Definitions.tex
\renewcommand{\a}{\mathbf{a}}

\newcommand{\h}{\mathbf{h}}

\newcommand{\x}{\mathbf{x}}

\newcommand{\0}{\mathbf{0}}

\renewcommand{\H}{\mathbf{H}}
\newcommand{\I}{\mathbf{I}}

\newcommand{\Sigmab}{\mathbf{\Sigma}}

\newcommand{\setK}{\mathcal{K}}

\newcommand{\setU}{\mathcal{U}}

\newcommand{\Compl}{\mbox{$\mathbb{C}$}}

\newcommand{\argmax}{\operatornamewithlimits{argmax}}

\newcommand{\Exp}{\mathbb{E}}

\newcommand{\herm}{\mathrm{H}}

\renewcommand{\Pr}{\mathbb{P}}

\newcommand{\tr}{\mathrm{tr}}
\newcommand{\tran}{\mathrm{T}}